\documentclass[aps,amsmath,reprint,amssymb,groupedaddress,longbibliography]{revtex4-1}
\usepackage{graphicx}
\usepackage{dcolumn}
\usepackage{bm}
\usepackage{bm}
\usepackage{gensymb}%
\usepackage[english]{babel}
\usepackage{txfonts}

\begin{document}

\title{Emergent scar lines in chaotic advection of passive directors} 

\author{Bardia Hejazi$^1$, Bernhard Mehlig$^2$, Greg A. Voth$^1$}
\email{gvoth@wesleyan.edu}
\affiliation{$^1$Department of Physics, Wesleyan University, Middletown, Connecticut 06459, USA}
\affiliation{$^2$Department of Physics, G\"{o}teborg University, 41296 Gothenburg, Sweden}

\date{\today}

\begin{abstract}

We examine the spatial field of orientations of slender fibers that are advected by a two-dimensional fluid flow.  The orientation field of these passive directors are important in a wide range of industrial and geophysical flows.  We introduce emergent scar lines as the dominant coherent structures in the orientation field of passive directors in chaotic flows. Previous work has identified the existence of scar lines where the orientation rotates by $\pi$ over short distances, but the lines that were identified disappeared as time progressed.  As a result, earlier work focused on topological singularities in the orientation field, which we find to play a negligible role at long times.  We use the standard map as a simple time-periodic two-dimensional flow that produces Lagrangian chaos.  This class of flows produces persistent patterns in passive scalar advection, and we find that a different kind of persistent pattern develops in the passive director orientation field.    We identify the mechanism by which emergent scar lines grow to dominate these patterns at long times in complex flows.  Emergent scar lines form where the recent stretching of the fluid element is perpendicular to earlier stretching.  Thus these scar lines can be labeled by their age, defined as the time since their stretching reached a maximum.

\end{abstract}

\maketitle

\section{Introduction}

When slender fibers are advected in a fluid flow, they become aligned by the flow \cite{Szeri:1991,Wilkinson2009,Parsa2011,Parsa2012,Ni2014,Guazzelli:2011,Voth2017} which produces dramatic effects including changes in material properties such as fluid rheology and scattering of electromagnetic waves.  These effects of fiber alignment appear in many applications including design of fiber suspension flows for the paper industry \cite{Olson1998,lundell2011}, prediction of the albedo of icy clouds \cite{Saunders1994, Saunders2001, Sherwood2006, Pinsky1998}, and controlling turbulent drag by adding fibers \cite{Dimitropoulos2005, Shaqfeh2005}.   Other applications include liquid crystals~\cite{deGennes:1995} and active nematics \cite{Keber2014,Giomi2015}.

The motion and alignment of small, slender fibers in fluid flow has many similarities to the advection of passive scalars such as the concentration of a dye.    This passive scalar problem has proven to be a rich area for scientific study~\cite{Ottino:1990,Warhaft:2003, Aref:2017}.  For passive scalars, the case of time periodic two-dimensional flows has been a source of many insights since it is the simplest case that produces Lagrangian chaos~\cite{Aref:1984}.   
A wide variety of mathematical tools have been developed for analyzing passive scalar advection~\cite{Aref:2017}.  Particularly relevant to fiber flows is analysis using finite time Lyapunov exponents (which quantify the stretching experienced by each infinitesimal fluid element) that has allowed insights from simple two-dimensional (2D) time-periodic flows to be extended to identification of Lagrangian coherent structures in complex flows~\cite{Haller2015}.     

The advection of small slender rigid fibers in fluid flow can be called the passive director problem.  Symmetric fibers are described by directors rather than vectors because the two anti-parallel orientations of the particle are equivalent.   The orientational degree of freedom of the director introduces physics that is not present in the passive scalar problem.  For passive directors, a flow produces non-trivial patterns in the orientation field  even for homogenous initial conditions leading to an entirely different class of problems~\cite{Szeri:1991}.   However, the basic phenomenology of the orientation field for passive director advection in 2D chaotic flow matches the passive scalar problem quite closely.   


Despite extensive study of the dynamics of fibers in fluid flows~\cite{Guazzelli:2011,Voth2017}, we still do not have a clear phenomenology of the fiber orientation field in chaotic and turbulent flows.  
Szeri \textit{et al.}~\cite{Szeri:1991,Szeri1992,Szeri1993a,Szeri1993b,Szeri:1994} analyzed the orientation dynamics of microstructured fluids in a framework applicable to rigid fibers as well as deformable microstructure such as polymers.  Their mathematical formalism describes cases where the fluid flow experienced by a particle is steady or periodic in time.  In these simple cases, they already found a rich range of phenomenology including chaotic dynamics of particle orientations.    Because these flows have integrable translational motion of particles, many interesting features of passive scalar advection do not yet occur.    
Two studies that explored the orientation field of passive directors in flows with chaotic fluid trajectories were performed by Wilkinson and co-workers~\cite{Wilkinson2009,Bezuglyy2010}.   They used a random flow in which they highlighted the existence of scar lines and topological singularities.    Another study by the same team~\cite{Wilkinson2010} extended the work of Szeri \textit{et al.} on flows with integrable translational trajectories.   Parsa~\textit{et al}~\cite{Parsa2011} performed an experimental study in which they measured the orientation of fibers in 2D chaotic and turbulent flows and identified how tools from continuum mechanics can be used to quantify fluid stretching and understand fiber orientations.  They only considered single fibers and not the spatial field of fiber orientation.

Another line of research has explored the alignment and curvature of fluid elements in chaotic and turbulent flows.  Fluid element orientation is closely related to passive director orientation, and so the curvature of fluid elements is related to the spatial gradient of the passive director orientation.    Pope \textit{et al.} used direct numerical simulations to analyze the curvature of material elements in turbulent flows~\cite{Pope1988,Pope1989}.  They found that the probability distribution of curvature approaches an asymptotic form while the mean square curvature diverges exponentially.  In 2D chaotic flows, the field of stretching and curvature of fluid elements has been analyzed to understand mixing~\cite{Liu1996,Giona1999,Thiffeault2004}.    Among other things, these studies explore a correlation between curvature and low stretching which was first observed in a study of model turbulent flow~\cite{Drummond1991}.  One particularly relevant result is the existence of asymptotic directionality in 2D time-periodic flows which causes fluid element orientations to approach a persistent pattern~\cite{Giona1999} similar to the persistent patterns observed in passive scalar advection~\cite{Pierrehumbert1994, Rothstein:1999}.   


There are several other problems where the spatial field of director orientations is studied.  Work on nematic liquid crystals has developed many of the tools to study these fields~\cite{deGennes:1995}.
Active nematics such as films of microtubules and molecular motors add additional dynamics to the nematic liquid crystal problem~\cite{Keber2014,Giomi2015}.   Studies of pattern formation in Rayleigh B\'enard convection also involve director fields formed by the orientation of convection rolls~\cite{Cross:1993,Egolf:1998}.   
 Studies of the polarization orientation in optics also involve a similar director field~\cite{Dennis:2008,Flossmann:2008}.  Recent work has shown the importance of director fields in the dynamics of cells~\cite{Kawaguchi2017,Saw2017,Duclos2017}.
These director fields all necessarily produce similar topological singularities, primarily those with a $\pm 1/2$ Poincar\'e index, which are like the core and delta ridge patterns first identified in the study of fingerprints.   We will see that the fluid passive director problem is an interesting case to contrast with the others.  It is a limiting case where the lack of interactions between directors leads to patterns with many features in common with patterns in the passive scalar problem.  

In this paper we identify the key coherent structures in fiber orientations in chaotic 2D flows and the mechanism by which they form.   We find that the topological singularities that have received extensive attention are not central to understanding fiber orientation fields.  Instead, it is scar lines~\cite{Wilkinson2009} that dominate the fiber orientation fields and we identify the mechanism that produces the emergent scar lines that dominate the field at long times.   
  
\begin{figure}[tb]
	\centering
	\includegraphics[width=0.5\textwidth]{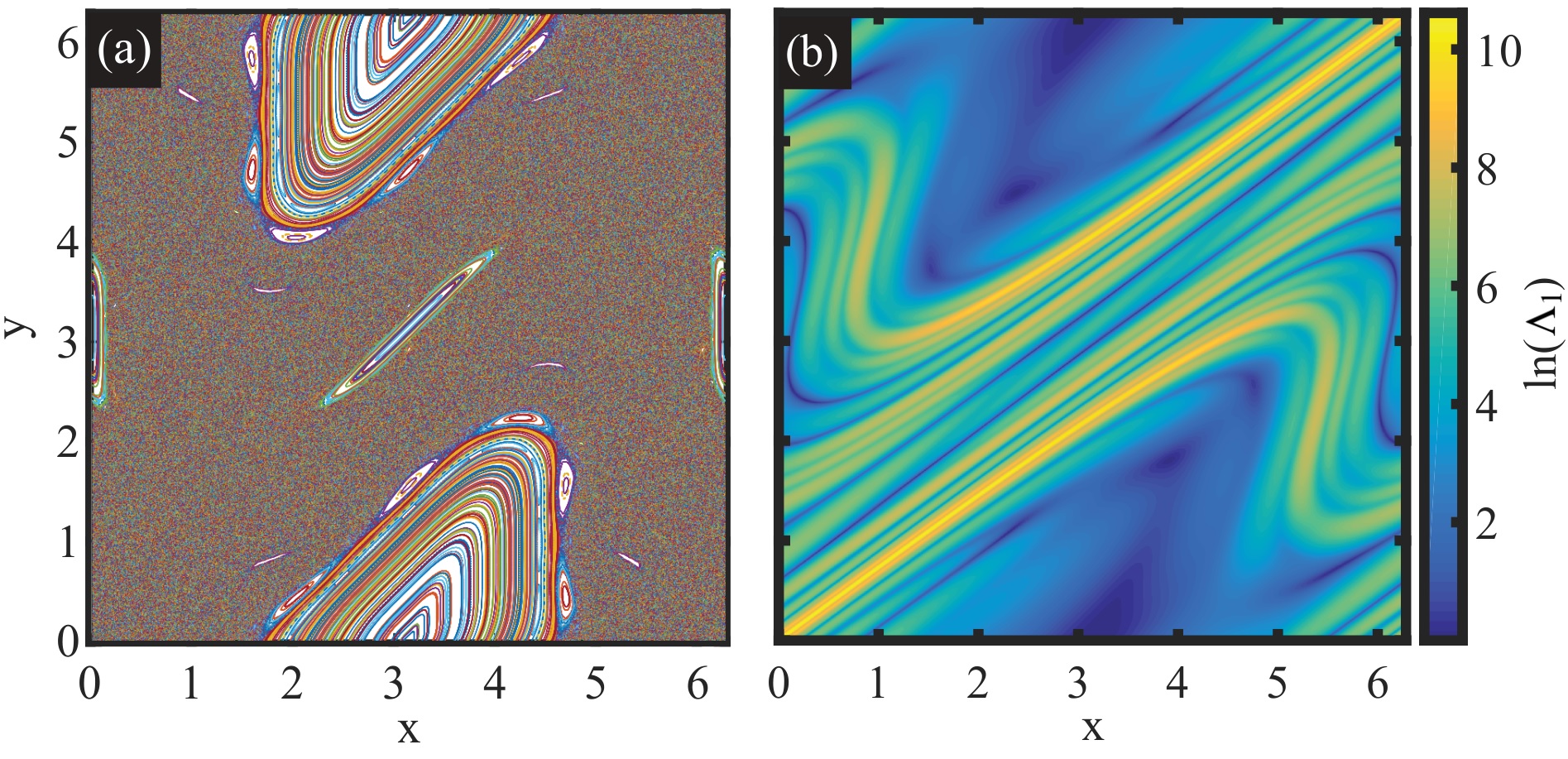}
	\caption{(a) Poincar\'{e} section of the standard map showing the regular regions and chaotic regions for $K=2$. (b) Stretching experienced by the fluid over four periods at the same value of $K$, where $\Lambda_1$ is the eigenvalue of the Cauchy-Green strain tensor defined in the Appendix.}
	\label{fig:Poincare_Stretch}
\end{figure}

\begin{figure*}[tb]
	\centering
	\includegraphics[width=0.7\textwidth]{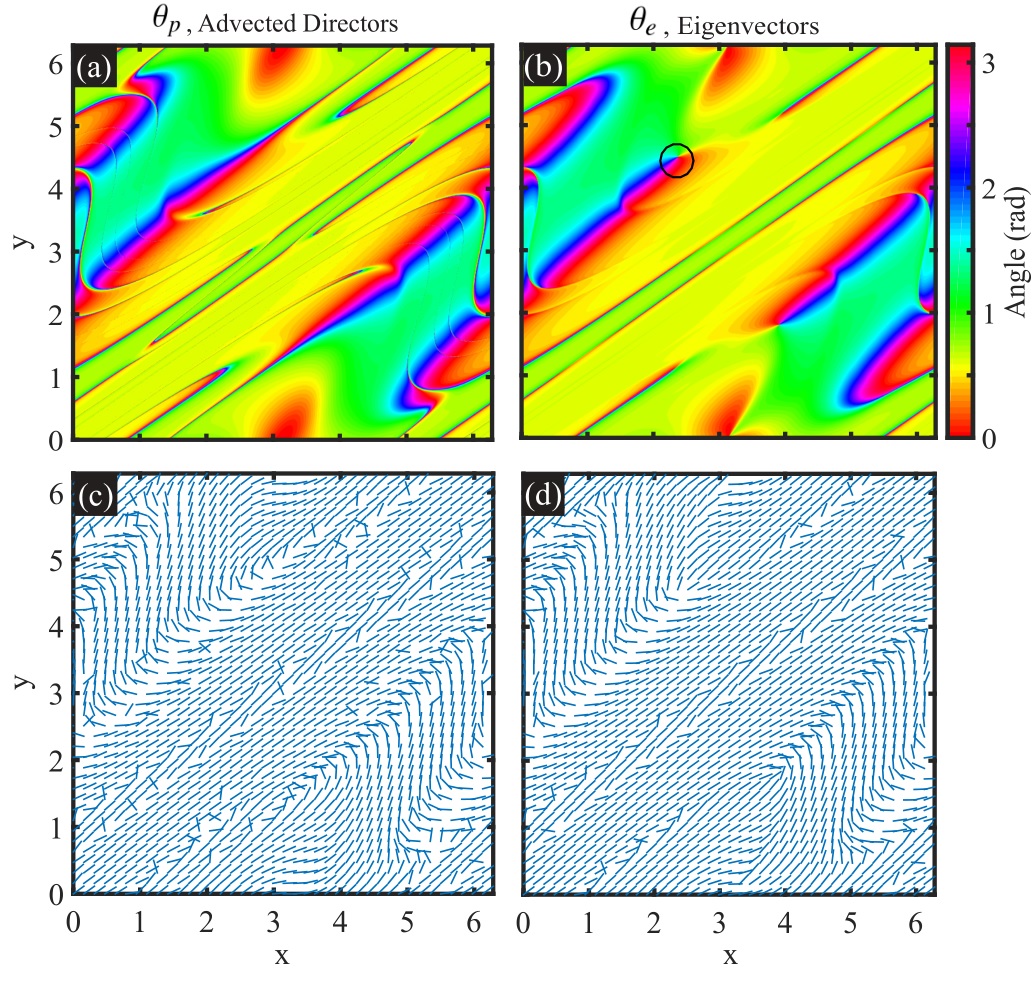}
	\caption{(a) Angle of the advected directors, $\theta_p$, of initially horizontal fibers. (b) Angle of the eigenvectors of the left Cauchy-Green strain tensor,  $\theta_e$. (c) and (d) Director representations of (a) and (b), respectively. All angles are measured with respect to horizontal.  Animations of all four of these figures are included as Supplemental Material~\cite{Supplemental_Material}.  (Here $K=2$ and $t=4$.)}
	\label{fig:Orientation_Quiver}
\end{figure*}

\section{Phenomenology of Passive Director Orientation in the Standard Map}

We study passive directors advected in the standard map, which is a convenient model for a two-dimensional fluid flow that exhibits Lagrangian chaos. The standard map is area preserving and invertible; it is defined as \cite{Lichtenberg1992}

\begin{eqnarray}
y_{t+1}=y_t+K\sin x_t
\label{eq:1}
\end{eqnarray}
\begin{align}
x_{t+1}=x_t+y_{t+1}
\label{eq:2}
\end{align}

where spatial coordinates $x$ and $y$ are periodic over $2\pi$, and $t$ is an integer that specifies the time measured in periods that the flow has been iterated.
This is often called the kicked rotor system and $q$ and $p$ are used instead of $x$ and $y$ for the phase space variables of the Hamiltonian dynamical system.   The standard map can be produced by a continuous flow field with the velocity in the first half of each period given by $\dot{x}=0$ and $\dot{y}=\frac{2K}{T} \sin x$ and in the second half of each period the velocity is given by $\dot{x}=\frac{2}{T} y$ and $\dot{y}=0$.  This flow alternates between a vertical Kolmogorov flow and a horizontal linear shear.   
A visualization of passive scalars in this flow is shown as an animation in the Supplemental Material~\cite{Supplemental_Material}. 
   
Figure~\ref{fig:Poincare_Stretch}(a) shows the Poincar\'{e}  section for the standard map for $K=2$ with the regular and chaotic regions clearly visible~\cite{Ott:2002}.  Figure~\ref{fig:Poincare_Stretch}(b) shows the field of fluid stretching often called the finite-time Lyapunov exponent field which is used to visualize Lagrangian coherent structures~\cite{Voth2002,Haller2015}.  

The orientation of a fiber advected in the flow defined by the standard map is
\begin{align}
\theta_{t+1}=\arctan \left(\frac{K\cos x_{t}+\tan \theta_{t}}{1+K\cos x_{t}+\tan \theta_{t}}\right).
\label{eq:3}
\end{align}
The orientation field of advected fibers can be defined in two different ways~\cite{Bezuglyy2010}.  Fibers with initial orientation field
$\hat{p}_{0}(\mathbf{r})$ evolve over time $t$ to a final orientation field $\hat{p}(\mathbf{r},t)$.  We will call this final orientation field the advected director field.  In two dimensions, this field is most easily represented by an orientation angle field $\theta_p(\mathbf{r},t)$.   Alternatively, each point can have a distribution of initial orientations, ${P}_{0}(\hat{p},\mathbf{r})$, which evolves under the flow to a distribution of final orientations, ${P}(\hat{p},\mathbf{r},t)$. The orientation field is then defined by the final preferred orientation of fibers at any point in space.  In the simplest case with uniformly distributed initial orientations, the preferred orientation can be obtained as the eigenvector of the left Cauchy-Green strain tensor (CGST) that corresponds to the maximum eigenvalue, which we denote by $\hat{e}_{L1}$. We will call this the eigenvector field and represent it by the orientation angle field $\theta_e(\mathbf{r},t)$.  The left Cauchy-Green Strain tensor is $\mathbf{C}^{(L)}=\mathbf{F}\mathbf{F}^{T}$
where $F_{ij}= \frac{\partial x_i}{\partial X_j}$ is the fluid deformation gradient~\cite{Malvern:1969,Parsa2011,Ni2014}, also referred to as the monodromy matrix \cite{Bezuglyy2010}.
An equivalent definition is to use an eigenvector of the tensor order parameter~\cite{deGennes:1995}.  The Appendix discusses the definitions and relationships between these quantities in more detail.     An important distinction between the advected director orientation field, $\theta_p(\mathbf{r},t)$ and the eigenvector orientation field, $\theta_e(\mathbf{r},t)$,  is that $\theta_p$  depends on a choice of initial orientation while $\theta_e$ does not. 

The two orientation fields, $\theta_p$ and $\theta_e$, are shown in Fig.~\ref{fig:Orientation_Quiver} at time $t=4$.   Both the fiber fields [Figs.~\ref{fig:Orientation_Quiver}(c) and \ref{fig:Orientation_Quiver}(d)] and higher resolution color maps of the orientation angles [Figs.~\ref{fig:Orientation_Quiver}(a) and \ref{fig:Orientation_Quiver}(b)] are shown.
The two different definitions of the fiber orientation field are quite similar, but there are some clear differences. For example, at $(x,y)=(2.37,4.48)$ in Fig.~\ref{fig:Orientation_Quiver}(b) we see that there is a pinwheel where the eigenvector orientation is not defined, a topological singularity, while at this point, the advected director field in Fig.~\ref{fig:Orientation_Quiver}(a) is smooth. We show in Sec.~\ref{sec:topo} that these fields have very different topological structure and yet as observed by Wilkinson \textit{et al.}~\cite{Wilkinson2009}, they converge toward the same field at long times in chaotic regions of the flow.    

\begin{figure}[tb]
	\centering
	\includegraphics[width=0.4\textwidth]{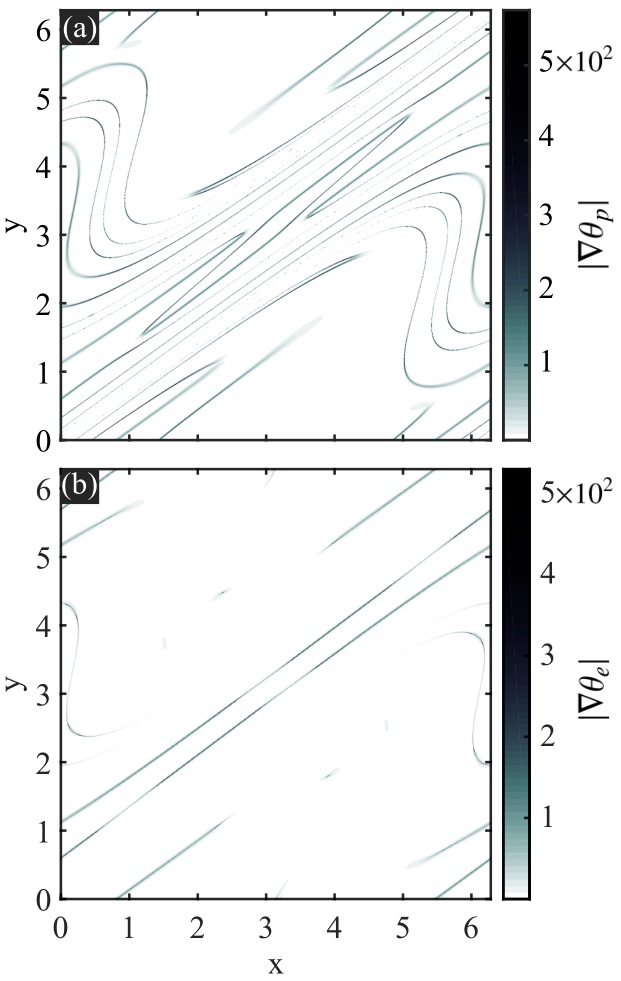}
	\caption{Gradient of the angle of the fiber orientation field after advection in the standard map for four periods. (a) Gradient of the advected director field starting with initially horizontal fibers.  (b) Gradient of the eigenvector field.  The large gradients lie on thin lines called scar lines~\protect{\cite{Wilkinson2009}}.
(Here $K=2$ and $t=4$.)}
	\label{fig:Gradient}
\end{figure}

An effective way to observe the dominant coherent structures in the orientation fields is to calculate the gradient of the fiber orientation, as shown in Fig.~\ref{fig:Gradient}. In both the advected director field [\ref{fig:Gradient}(a)] and the eigenvector field [\ref{fig:Gradient}(b)], the dominant features are thin lines over which the orientation changes by $\pi$ over a very short distance.  These have been called scar lines by Wilkinson \textit{et al.}~\cite{Wilkinson2009}.    

The basic mechanism for formation of a scar line is simple.  When fluid is stretched by the flow, fibers rotate toward alignment with the stretching direction.  However, some fibers that are initially perpendicular to the stretching direction will not align.   The set of points with initial orientations that are exactly perpendicular to the stretching direction fall on lines.  In chaotic regions of the flow where stretching increases exponentially in time, the width of the perpendicular region is shrinking exponentially in time, causing the orientation field to rotate by $\pi$ across very short distances.     

It is not immediately obvious how the mechanism in the preceding paragraph creates scar lines in the eigenvector field.   Wilkinson \textit{et al.}  briefly identify type 2 scar lines as lines that form where rotational regions with complex eigenvalues of the deformation gradient (or monodromy) matrix have repeatedly been stretched and folded so that they become very narrow.   Complex eigenvalues appear where the trace of the monodromy matrix is between $-2$ and 2.   We find that the scar lines in the eigenvector field appear in regions of low stretching that are often associated with these stretched rotational regions.   However, the mechanism we observe for the formation of scar lines in the eigenvector field does not clearly match the definition of type 2 scar lines by Wilkinson \textit{et al.}.   In Sec.~\ref{sec:emergent} we show how the scar lines in the eigenvector field emerge as the result of an initial stretching that creates a preferred orientation.  When the later stretching experienced by that fluid element is perpendicular to the initial stretching, a scar line is created.  We call these emergent scar lines and find that they dominate the orientation field of both advected directors and stretching eigenvectors at large times.  The reversal of stretching results in scar lines being associated with regions of low stretching, connecting with earlier observations that curvature of fluid elements preferentially occur in regions of low stretching~\cite{Drummond1991,Liu1996,Thiffeault2004}.  The mechanism for creation of emergent scar lines is similar to the mechanism for creation of type 1 scar lines except that the stretching over an initial time interval replaces the initial fiber orientation.      At large times, type 1 scar lines become unobservably thin so that emergent scar lines that have been formed in the recent past dominate the observed orientation fields. In Sec.~\ref{sec:age} we show how these emergent scar lines can be labeled by the time since their creation.  
 
\begin{figure}[tb]
	\centering
	\includegraphics[width=0.5\textwidth]{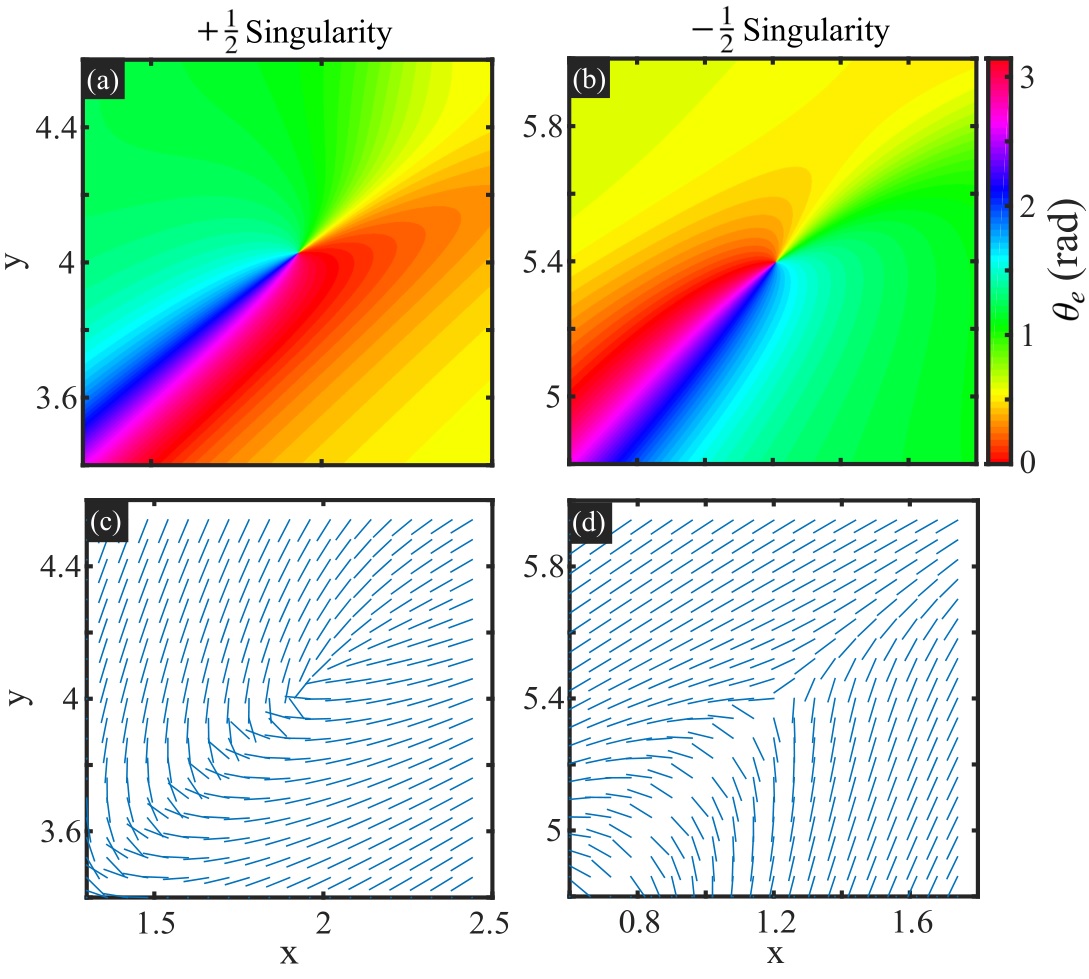}
	\caption{The two different types of singularities. (a) and (b) Angle of $\hat{e}_{L1}$ with respect to the horizontal. (c) and (d) Director representation of the eigenvector field. (a) and (c) show a singularity with a Poincar\'e index of $+\frac{1}{2}$ and (b) and (d) show a singularity with a Poincar\'e index of $-\frac{1}{2}$.  To determine the Poincar\'e index, circle the singularity of (c) in the clockwise direction.  Around the circle, the orientation of $\theta_e$ rotates by $\pi$ in the clockwise direction, giving a Poincar\'{e} index of $+\frac{1}{2}$. (Here $K=2$ and $t=2$.)}
	\label{fig:Singularity_types}
\end{figure}

\begin{figure*}[tb]
	\centering
	\includegraphics[width=1\textwidth]{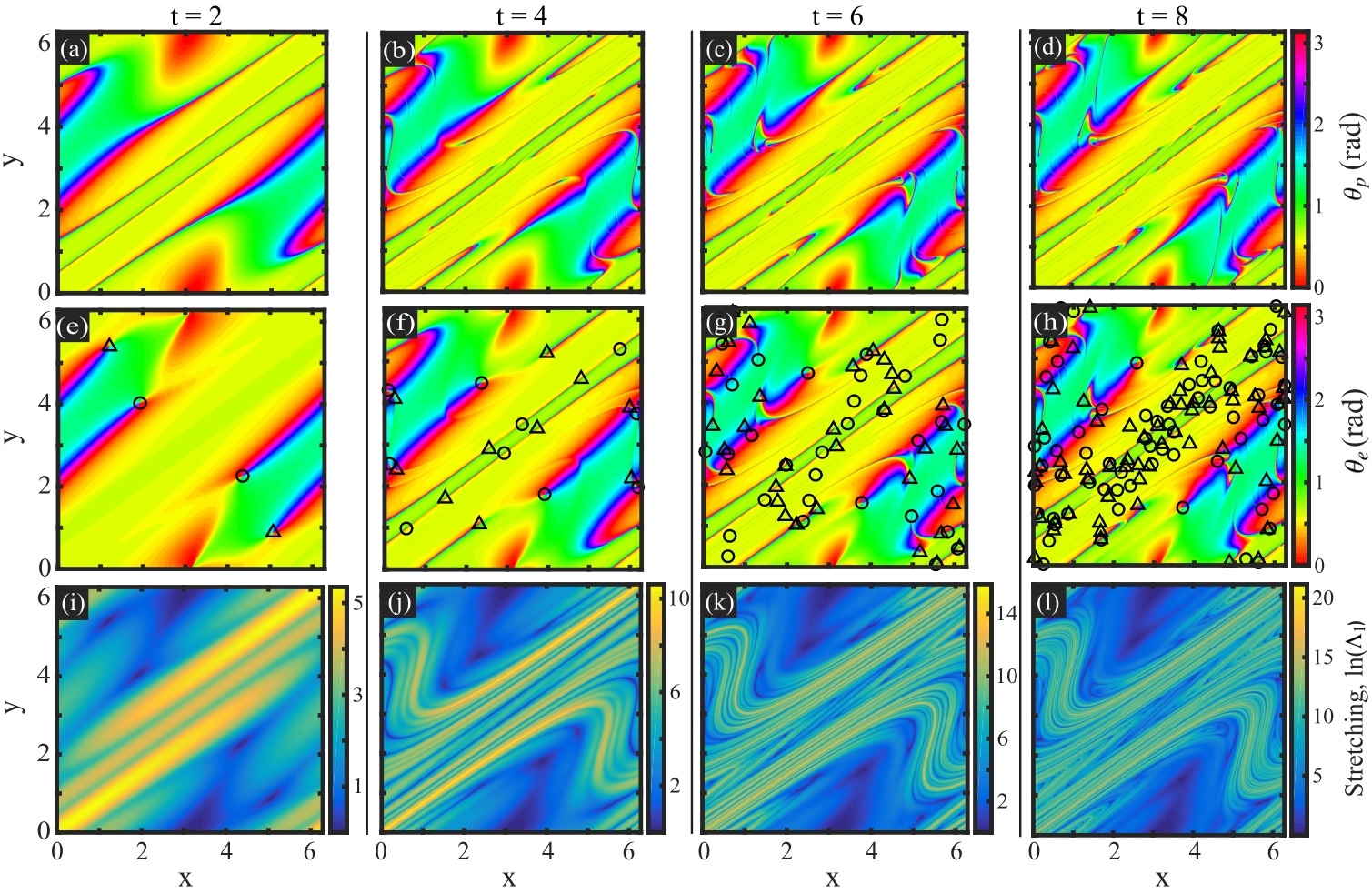}
	\caption{(a)-(d) Angle of advected directors, $\theta_p$, measured from horizontal over the time interval $t=2$, $4$, $6$, and $8$. (e)-(h) Angle of stretching eigenvectors, $\theta_e$, over the same time intervals with singularities marked: $\medcirc$, $+\frac{1}{2}$ singularities; $\bigtriangleup$, $-\frac{1}{2}$ singularities. (i)-(l)  Stretching experienced by the fluid for the same time intervals.  Animations of these three fields are provided as Supplemental Material~\cite{Supplemental_Material}. (Here $K=2$.)}
	\label{fig:12figs_time_evolution}
\end{figure*}
\section{Topological Singularities in Passive Director Orientation Fields}
 \label{sec:topo}

A major focus in previous work on the evolution of director fields has been the development of topological singularities or topological defects~\cite{Cross:1993,Egolf:1998,Bezuglyy2010,Keber2014,Giomi2015,deGennes:1995,Wilkinson2010}. To conserve the total topological charge of the field, singularities must always form in pairs of opposite Poincar\'{e} indices. The two types of singularities that form in director fields are shown in Fig.~\ref{fig:Singularity_types}. Figures~\ref{fig:Singularity_types} (a) and \ref{fig:Singularity_types}(c) show a singularity that has a Poincar\'{e} index of $+\frac{1}{2}$; and Fig.~\ref{fig:Singularity_types} (b) and (d) show a singularity which has a Poincar\'{e} index of $-\frac{1}{2}$.   The Poincar\'{e} index is defined as the number of multiples of $2\pi$ in which the director orientation changes as we move around a closed loop.  
These singularities are given different names by different communities.  The terminology from fingerprint analysis is core for $+\frac{1}{2}$ and delta for $-\frac{1}{2}$.  They are also called wedge and trisector~\cite{Karrasch:2014}, and in the study of optical polarization fields there are similar singularities called star and lemon~\cite{Flossmann:2008}.  Recent work on patterns in cell populations has used `comet-like' for $+\frac{1}{2}$ singularities~\cite{Kawaguchi2017,Saw2017,Duclos2017}.

Figure~\ref{fig:12figs_time_evolution} shows how the orientation fields and the stretching field develop from $\Delta t=2$ to $8$. As time progresses, the structure of the advected director field $\theta_p$ and the eigenvector field  $\theta_e$ become more alike.  However,  certain regions are still different.  The differences occur in regions of low stretching that are either in the elliptic islands of the flow (see Fig.~\ref{fig:Poincare_Stretch}) or within the thin lines where the stretching is small.

In Fig.~\ref{fig:12figs_time_evolution}(a)-\ref{fig:12figs_time_evolution}(h) the topological singularities are marked with circles and triangles.   The eigenvector field continuously develops new topological singularities while the advected director field always remains free of them. This difference occurs because the directors start as a smooth field and are advected by a smooth flow, so it is not possible for topological singularities to form~\cite{Wilkinson2009}.  In contrast, the stretching eigenvector field nucleates pairs of singularities at points where the stretching is zero.
An animation showing the generation of singularities in the field of $\theta_e$ is provided as Supplemental Material~\cite{Supplemental_Material}.

\begin{figure}[tb]
	\centering
	\includegraphics[width=0.35\textwidth]{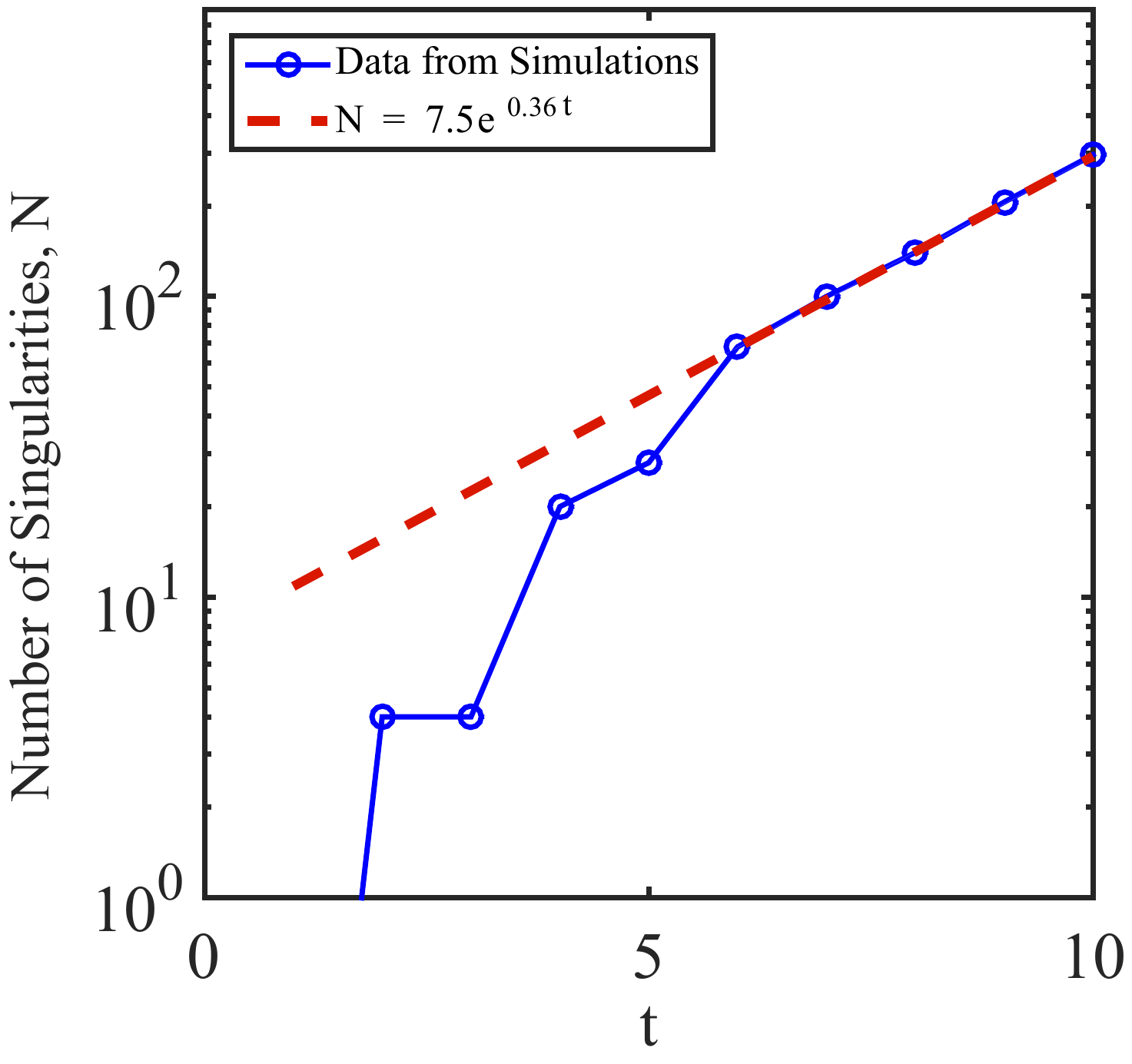}
	\caption{Number of singularities $N$ as a function of time. At later times the number of singularities increases exponentially. (Here $K=2$.)}
	\label{fig:Number_of_singularities}
\end{figure}

Figure~\ref{fig:Number_of_singularities} shows a plot of the number of singularities in the stretching eigenvector field at each period, $N$. 
  After an initial transient, the number of singularities grows exponentially.  A least squares fit to $t > 5$ gives $N=7.5 \, e^{0.36t}$.  
The exponential can be understood as the result of a process similar to a baker's map where the thin lines of low stretching are folded on themselves multiple times.   New singularities are nucleated in these low stretching regions, so the number of new zeros in the stretching at each period is proportional to the current number of zeros.  

The number of singularities were calculated computationally by performing a non-linear search for minima where the stretching field is near unity.  At these points that have not been stretched, the eigenvector field has no preferred orientation, allowing a singularity in the field of $\theta_e$.   Serra and Haller~\cite{Serra:2017} have shown that care is required because integrating trajectories in noisy or intermittent velocity fields can create artificial singularities in the stretching eigenvector field.  We have not observed any artificial singularities, likely because of the simple analytic expressions for the standard map.  The number of singularities is increasing rapidly with time, while the spatial extent over which the stretching is near unity becomes very small.  As a result, it becomes increasingly difficult to find all the singularities as time progresses.  We will see that the shrinking of the size of the region affected by each singularity allows the exponentially increasing number of singularities to become less and less important to the orientation field as time progresses.

By comparing the number of singularities found in separate computations with a very large number of initial guesses, we confirmed that we were able to find all singularities up to $t=10$, but by $t=12$ it is clear that we were missing a significant number and so we only report data up to $t=10$. To characterize the topological charge (Poincar\'{e} index) of the singularities we move around each singularity in a small loop (500 points around a circle of radius $10^{-5}$) and calculate the change in orientation of fibers around that loop.

Since the flow is defined by a simple analytic map, we can calculate the positions at which singularities form over the first few periods.   At a singularity, the stretching is zero and so the deformation gradient $\mathbf{F}$ represents a pure rotation.  Here we analytically calculate the positions of singularities that appear over the time range $t=1.5$$-$$2$.  The position of a fluid element initially at $(x_0,y_0)$ advected over time $t=\frac{3}{2}+ \frac{\varepsilon}{2}$, where $0 \leq \varepsilon \leq 1$, is

\begin{eqnarray}
\begin{pmatrix} x_{1+\varepsilon}\\ \\y_{2} \end{pmatrix}=
\begin{pmatrix}
x_{0}+(1+\varepsilon)(y_{0}+K\sin x_{0}) \\
+\varepsilon K\sin(x_{0}+y_{0}+K\sin x_{0})
\\ 
\\
 y_{0}+K\sin x_{0}\\+K\sin(x_{0}+y_{0}+K\sin x_{0})
\end{pmatrix}
\label{eq:xy32}
\end{eqnarray}


Singularities can exist at points where the deformation gradient $F$ calculated from from Eq.~\ref{eq:xy32} is a pure rotation so it satisfies $F_{11}=F_{22}$ and $F_{12}=-F_{21}$. 

The points that satisfy this condition are:
\begin{eqnarray}
\begin{cases}
x_{0}=2\pi m +\cos^{-1}\left(\frac{-1\pm \varepsilon}{K}\right) \\
\\
y_{0}=2\pi l \pm\cos^{-1}\left(\frac{(1+\varepsilon)\cos x_{0}}{1-\varepsilon-2\varepsilon \cos x_{0}}\right)-x_{0}- 2\sin x_{0} 
\end{cases}
\label{eq:sing1}
\end{eqnarray}

and
\begin{eqnarray}
\begin{cases}
x_{0}=2\pi m -\cos^{-1}\left(\frac{-1\pm \varepsilon}{K}\right) \\
\\
y_{0}=2\pi l \pm\cos^{-1}\left(\frac{(1+\varepsilon)\cos x_{0}}{1-\varepsilon-2 \varepsilon\cos x_{0}}\right)-x_{0}+2\sin x_{0} 
\end{cases}
\label{eq:sing2}
\end{eqnarray}
where $l$ and $m$ are integers. 
Note that these are initial coordinates $(x_0,y_0)$.  The position of singularities are the final coordinates, $(x_{1+\varepsilon},y_{2})$, which are obtained by inserting the values we find for $(x_0,y_0)$ into Eq.~(\ref{eq:xy32}).  For $\varepsilon=0$ and 1, the positions of the singularities are periodic over $2 \pi$.    For $0< \varepsilon < 1$, we choose $m$ and $l$ so that the final singularity positions lie within $[0, 2 \pi]$.

These analytical calculations for the position of singularities agree exactly with the positions found computationally in Fig.~\ref{fig:12figs_time_evolution}(e).  
Figure~\ref{fig:12figs_time_evolution}(e) shows only four singularities, because out of the eight singularities in Eqs.(\ref{eq:sing1}) and (\ref{eq:sing2}), four come together in pairs and annihilate at the large elliptical island leaving four singularities after two periods while conserving the total topological charge throughout the process.  The dynamics of the generation and annihilation of singularities in the stretching eigenvector field are shown in an animation that has been provided as Supplemental Material~\cite{Supplemental_Material}. 

\begin{figure}[tb]
	\centering
	\includegraphics[width=0.5\textwidth]{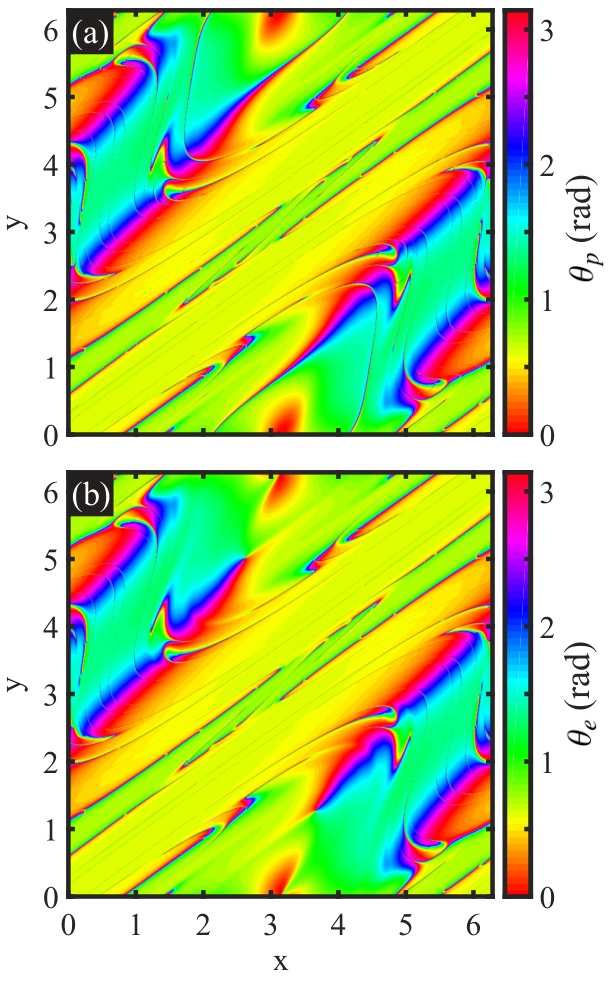}
	\caption{Orientation fields at longer times that show the persistent pattern that appears in the chaotic region of both fields.  (a) Orientation field of the advected directors $\theta_p$. (b) Orientation field of the stretching eigenvectors $\theta_e$. (Here $K=2$ and $t=10$.)}
	\label{fig:long_time_limit}
\end{figure}

\section{Scar Lines in Passive Director Orientation Fields}

The orientation fields of advected directors and stretching eigenvectors become very similar to one another and approach a stationary state in the long time limit, as is evident in Fig.~\ref{fig:12figs_time_evolution}.    In Fig.~\ref{fig:long_time_limit} these fields are shown at $t=10$.  In the chaotic regions of the flow, the two fields appear to become almost identical in the long-time limit, reflecting the persistent pattern~\cite{Pierrehumbert1994, Rothstein:1999} or asymptotic directionality~\cite{Giona1999} that has been observed in other work on time-periodic 2D flows.  However, the fields are also diverging in topology since there is an exponentially increasing number of singularities in the field of stretching eigenvectors, $\theta_e$.
Since the two fields converge throughout almost the entire chaotic region and yet have completely different topologies, the key coherent structures in these fields are apparently not the topological singularities.   Figure~\ref{fig:Gradient} suggests that instead, the key coherent structures are thin lines across which the fiber orientation rotates by $\pi$.   Wilkinson \textit{et al}~\cite{Wilkinson2009} have called these structures scar lines. In the long-time limit the topological singularities only affect an infinitesimally small region of the stretching eigenvector orientation field and are screened by the scar lines that come to be the dominant features of the field.

\begin{figure}[tb]
	\centering
	\includegraphics[width=0.45\textwidth]{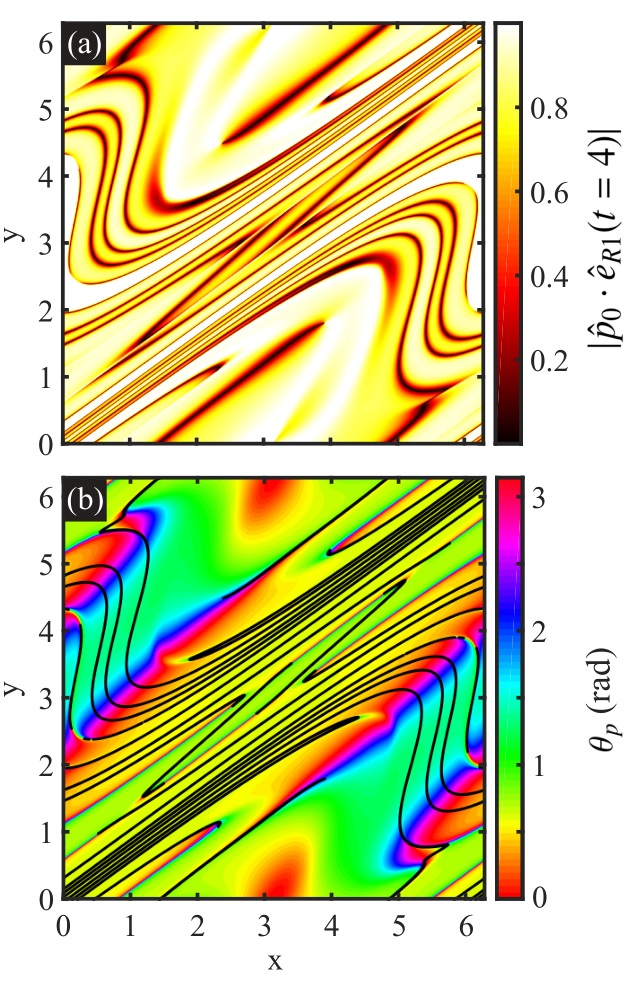}
	\caption{Locations of type 1 scar lines where the initial fiber orientation is perpendicular to the direction the fluid will be stretched.  (a) Dot product of the initial orientation with the maximum eigenvector of the right CGST $\hat{p}_0 \cdot \hat{e}_{R1}$.  (b) Locations where $\hat{p}_0 \cdot \hat{e}_{R1}< 0.03$ superimposed on the stretching eigenvector orientation field. (Here $K=2$ and $t=4$.)}
	\label{fig:Type1}
\end{figure}

\subsection{Type 1 Scar Lines}

Type 1 scar lines form in the advected director field at points where the initial fiber orientation is perpendicular to the direction that the fluid will be stretched.  The right Cauchy-Green strain tensor,  $\mathbf{C_R}=\mathbf{F^T} \mathbf{F}$ (see the Appendix), has eigenvectors that indicate the directions of stretching in initial particle coordinates~\cite{Malvern:1969,Parsa2011}.   In Fig.~\ref{fig:Type1}(a), we show the dot product of initial fiber orientation with $\hat{e}_{R1}$, the extensional eigenvector of the right Cauchy-Green strain tensor.  Type 1 scar lines form where this dot product is zero.  Figure.~\ref{fig:Type1}(b) shows the locations where this dot product is less than 0.03 superimposed on the advected director field.  The match with the locations of most of the scar lines is very good.  Some of the locations with a zero dot product do not initially appear to be scar lines, but at higher resolution it becomes clear that the scar line had simply become too thin to see at the 2000 $\times$ 2000 resolution of the plotted orientation field.   This healing of type 1 scar lines was identified by Wilkinson \textit{et al.}~\cite{Wilkinson2009}, and here we see that in this flow it only takes four periods for many of the type 1 scar lines to become too thin to be observed.  There are other points at which scar lines appear, but the zero dot product condition from Fig.~\ref{fig:Type1}(a) is not met.  These will be the topic of Sec.~\ref{sec:emergent}.

\begin{figure}[tb]
	\centering
	\includegraphics[width=0.5\textwidth]{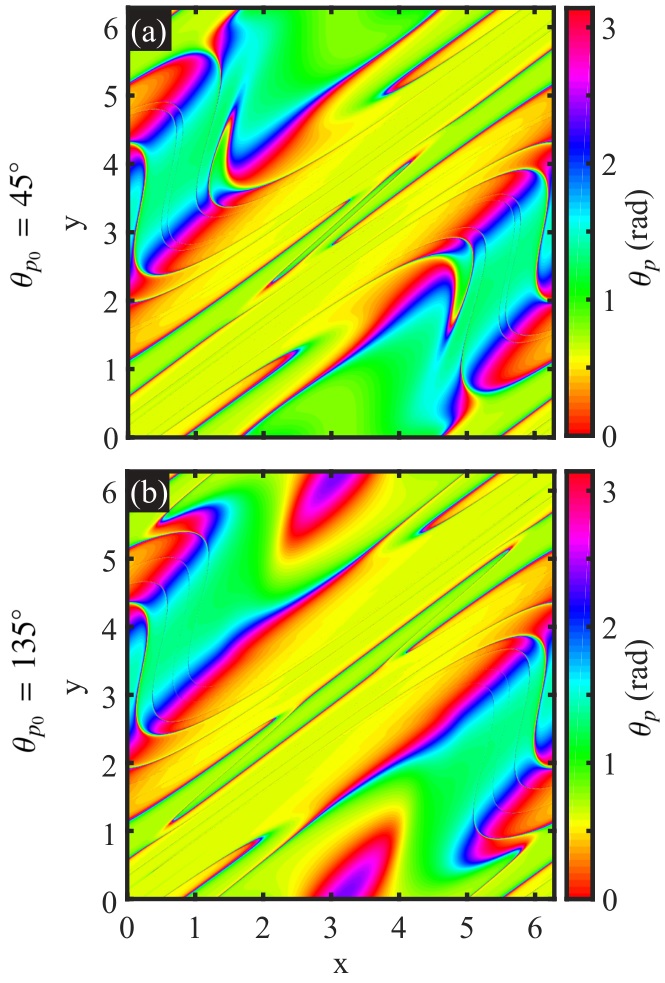}
	\caption{Sensitivity of the advected directors and type 1 scar lines to initial orientation. (a) Orientation field $\theta_p$ of an initially uniform grid of directors with an angle of $+45\degree$. (b) Orientation field $\theta_p$ of an initially uniform grid of directors with an angle of $+135\degree$. (Here $K=2$ and $t=4$.)}
	\label{fig:Initial_angle}
\end{figure}

Due to their production mechanism, the type 1 scar lines are sensitive to initial orientations of the advected directors. In Fig.~\ref{fig:Initial_angle} the orientation field for the advected directors, $\theta_p$, is shown for two different uniform initial orientation angles. It can be seen from this figure that although the two have differences caused by their initial orientations, the structure and topology of the field remains mostly the same and are mainly dominated by another type of scar line that is similar in both fields and independent of initial orientations. We identify these scar lines as emergent scar lines.

\subsection{Emergent Scar Lines}
\label{sec:emergent}
In this section we study the mechanism that creates scar lines that are independent of initial conditions and the distant past history of the flow.   We find that the dominant structures in the fields of both $\theta_e$ and  $\theta_p$ are emergent scar lines that develop when the recent stretching of a fluid element is orthogonal to the stretching it experienced earlier.  We have established that after sufficient time, there is a persistent pattern in the fiber orientation field.  This orientation pattern can be thought of as the initial orientation field for the flow over the next time interval.  The structures with a large gradient in the fiber orientation field occur where the stretching over the next time interval is orthogonal to the orientation produced by the stretching of the previous time interval. Quantitatively, a scar line will emerge where the stretching over some initial interval, which is defined by $\hat{e}_{L1}$, is perpendicular to the stretching the fluid element will experience in a future interval $\hat{e}_{R1}$.  Since we are using a periodic flow, we can calculate these stretching directions at any time and identify the locations where the initial stretching is orthogonal to the later stretching.

\begin{figure}[tb]
	\centering
	\includegraphics[width=0.45\textwidth]{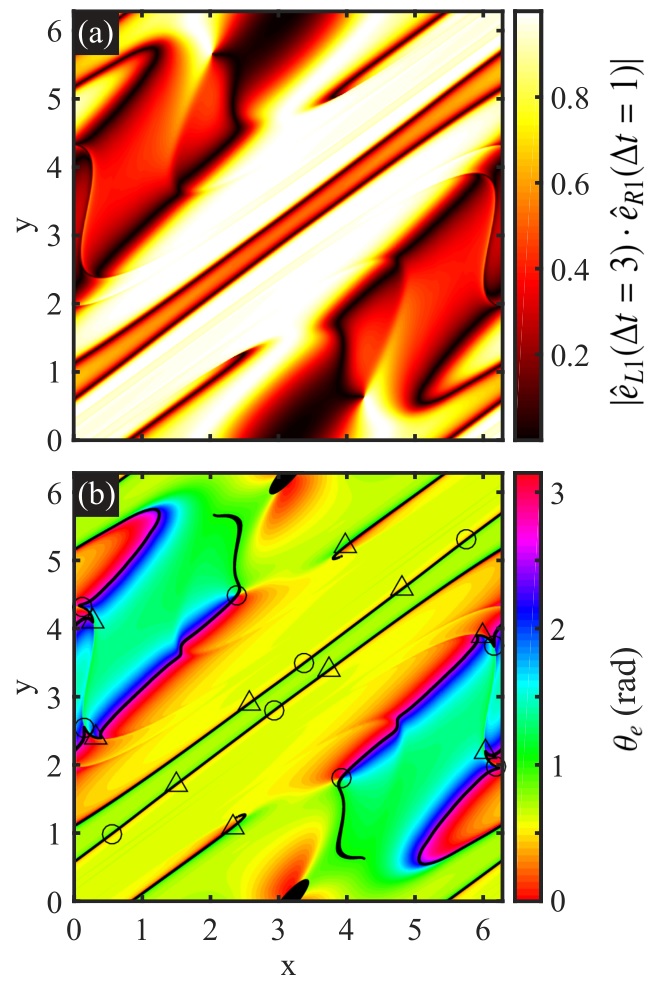}
	\caption{Locations of emergent scar lines.  (a) Dot product of the maximum eigenvector of the left CGST for three periods and the maximum eigenvector of the right CGST for one period $\hat{e}_{L1}(\Delta t=3) \cdot \hat{e}_{R1}(\Delta t=1)$. (b) Locations where $\hat{e}_{L1}(\Delta t=3) \cdot \hat{e}_{R1}(\Delta t=1)< 0.03$ superimposed on the stretching eigenvector orientation field.   (Here $K=2$ and $t=4$.)}
	\label{fig:Emergent}
\end{figure}

Figure~\ref{fig:Emergent}  verifies this mechanism for the formation of emergent scar lines. Figure~\ref{fig:Emergent}(a) shows the dot product of the stretching that the fluid has experienced over three periods  $\hat{e}_{L1}(\Delta t=3)$ with the stretching that the fluid will experience for the one remaining period $\hat{e}_{R1}(\Delta t=1)$.  The dot product of these vectors should be zero at the locations of the emergent scar lines for $\Delta t=4$, and Fig.~\ref{fig:Emergent}(b) shows this condition, and marks precisely the locations of the scar lines in the stretching eigenvector field in the chaotic region of the flow at $t=4$.  In the large regular islands, the stretching is small and some spurious points meet the condition of the dot product without developing scar lines. 

\begin{figure}[tb]
	\centering
	\includegraphics[width=0.45\textwidth]{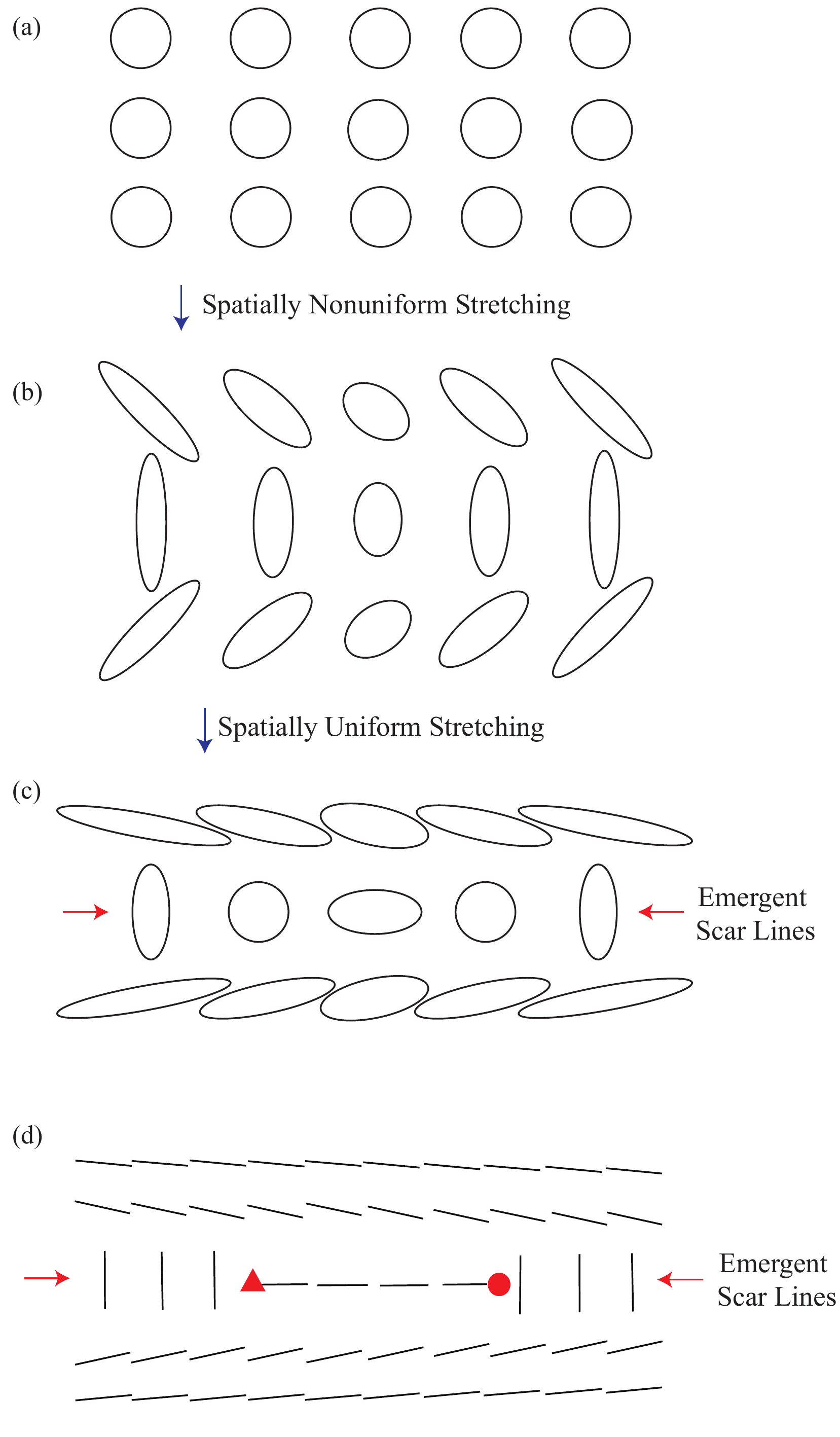}
	\caption{Simple schematic flow that generates singularities and emergent scar lines. (a) Uniform circular fluid elements.  (b) Fluid elements after deformation by a spatially non-uniform flow field. (c) Fluid elements after a uniform pure strain flow with horizontal extension direction. Emergent scar lines form in the second row of fluid elements where the uniform stretching is orthogonal to the initial non-uniform stretching. (d) Director representation of the stretching eigenvector after the two stretching steps.   The circle represents a $+\frac{1}{2}$ singularity and the triangle a $-\frac{1}{2}$ singularity.}
	\label{fig:Sketch}
\end{figure}

Figure~\ref{fig:Sketch} demonstrates the mechanism by which emergent scar lines and singularities are generated in the eigenvector field. This simple schematic flow consists of two steps.  First, the initially circular fluid elements experience a non-uniform flow field that stretches and rotates them to the arrangement shown in Fig.~\ref{fig:Sketch}(b).  Second, the fluid elements experience a uniform pure strain flow with a horizontal extension direction that results in the shapes in Fig.~\ref{fig:Sketch}(c). In this process there are points where the initial stretching experienced by some fluid elements is exactly canceled by later stretching. These points lie on the emergent scar lines where the  direction of the previous stretching is perpendicular to the later stretching. These points that experience no net stretching are the singularities of the stretching eigenvector field. Figure~\ref{fig:Sketch}(d) shows a director representation of the final configuration of the stretching direction with the singularities marked.   

\begin{figure}[tb]
	\centering
	\includegraphics[width=0.48\textwidth]{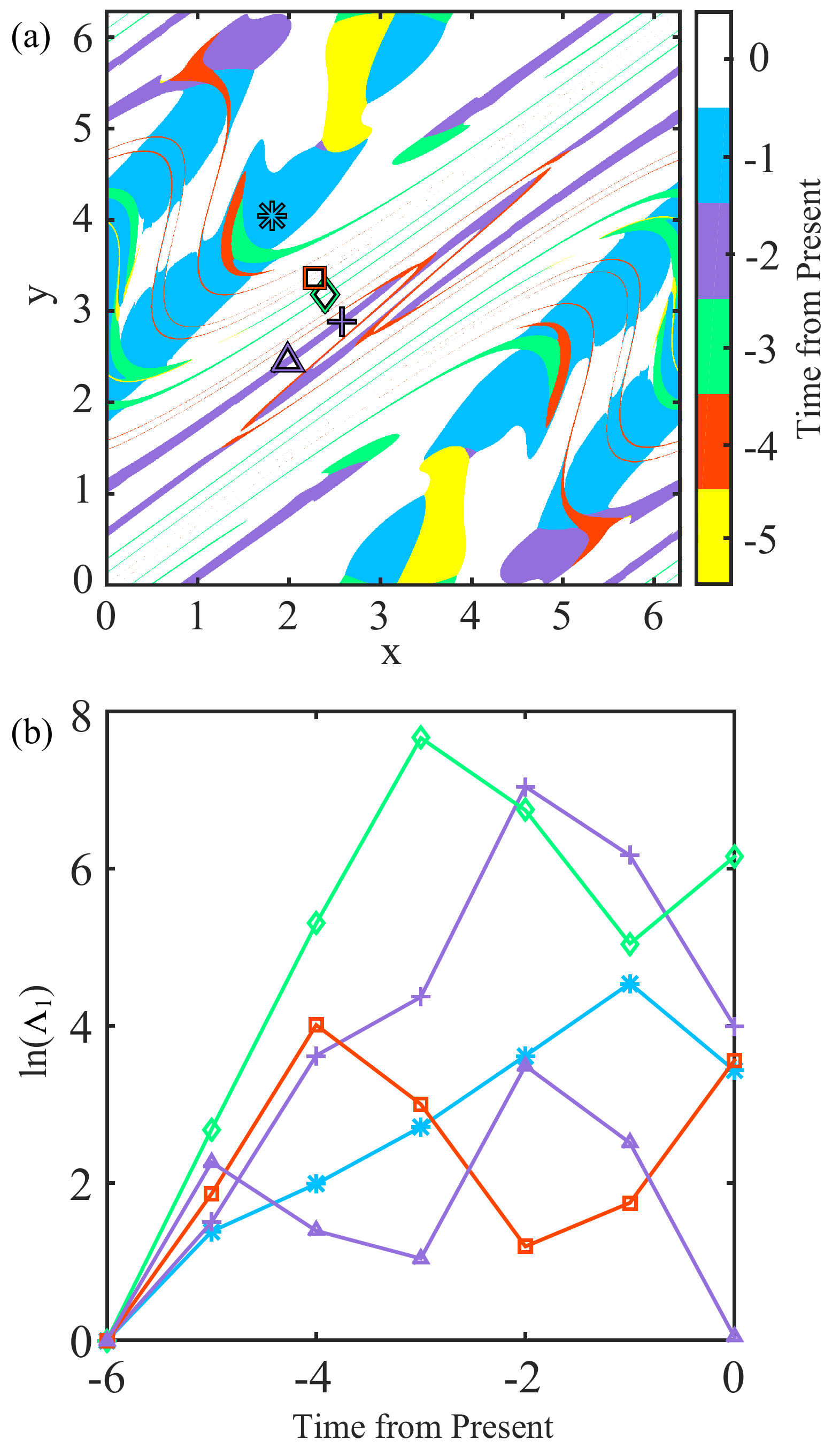}
	\caption{Quantifying the age of emergent scar lines. (a) Field showing the time at which each fluid element experienced maximum stretching, which we use to define the age of emergent scar lines.  Most of the field is white, indicating that maximum stretching occurs at the current time, but several generations of scar lines appear with maximum stretching before the present time. (b) Stretching as a function of time for points that lie on emergent scar lines of different ages. $\ast$, age $1$ emergent scar line; $+$, age $2$; $\diamond$, age $3$; $\square$, age $4$; $\bigtriangleup$, $-\frac{1}{2}$ singularity on emergent scar line of age $2$. (Here $K=2$ and $t=6$.)}
	\label{fig:Gen_map}
\end{figure}
\subsection{Age of Emergent Scar Lines}
\label{sec:age}

An emergent scar line forms when earlier stretching is reversed by later stretching.   The time that divides earlier from later is different for different scar lines.  Here we show how each emergent scar line can be labeled with its age defined as the time since its maximum stretching occurred and reversal of stretching began. 
 Since stretching increases exponentially and scar line thickness decreases exponentially, there is often an age beyond which the emergent scar lines become so thin that they are below the resolution of interest.  In experiments, there will be resolution limits and translational diffusion that will remove old scar lines that have become sufficiently thin.  The scar lines that are young enough to remain visible at the resolution of interest dominate the passive director orientation field.

    In Fig.~\ref{fig:Gen_map} we show a field indicating the time when the maximum stretching occurred for each fluid element.  Most of the field is white, indicating that the maximum stretching is at the current time.  The thick blue region indicates the points where the maximum stretching occurred one period ago.  Purple indicates points whose maximum stretching was two periods ago marking an emergent scar line with an age of two periods.  In this field created with an integration over six periods, we observe scar lines with ages up to four periods.   (There is a part of the elliptic island that had its maximum stretching five periods ago, but without exponential stretching this does not form a scar line.)    Figure~\ref{fig:Gen_map}(b) plots the history of the stretching at the five points marked in Fig.~\ref{fig:Gen_map}(a).  The purple curve marked with a triangle is a topological singularity where the stretching goes to zero at time $t=6$.  It lies on an emergent scar line with age 2, labeled purple, since its maximum stretching was two periods before the present.  The other four points are chosen to lie on scar lines with ages of one to four periods.  

Once an emergent scar line becomes well defined so that its width is much less than the correlation length of the velocity gradients in the flow, it will be advected by the flow without being removed.   Later stretching will rotate the line and decrease its width by compressing the rotation by $\pi$ to a narrower region.  The scar line remains where orientation was perpendicular to the later stretching, and since the scar line has orientations across the full range 0 to $\pi$; this is guaranteed to occur in some region within the scar line.   

A fluid element can have multiple maxima in its stretching history so that multiple ages can be assigned to it.  The age 4 point shown in red in Fig.~\ref{fig:Gen_map}(b) has later stretching that has almost surpassed the maximum from four periods ago.  Some of these points are simply at the edge of the scar line and at longer times will cease to be part of the scar line.    However, some other points have had the recent stretching become larger than the earlier stretching.  This creates topological singularities by the process shown in the simple model in Fig.~\ref{fig:Sketch}, and the dynamics of scar lines near these topological singularities is a topic that needs additional study.   At long times, the fraction of the chaotic region where the stretching is small enough that recent stretching can overcome earlier stretching becomes very small.  So an exponentially increasing number of topological singularities are occurring in a shrinking fraction of the chaotic domain in a way that allows the overall structure of the orientation field to be independent of what happens in these regions near the singularities.  

%
%
%
%
%
%
%

\section{Conclusions}

When fibers are advected in two-dimensional flows with exponential stretching of material elements, the primary coherent structures in the fiber orientation field are scar lines over which the fiber orientation rotates by $\pi$ over short distances.  We have discovered that recently formed scar lines dominate the observed orientation fields. A scar line emerges in regions where the recent stretching is perpendicular to the earlier stretching of that fluid element. These emergent scar lines can be labeled by their age, defined as the time at which their stretching reached a maximum.

It is important to distinguish two different ways to quantify the director orientation field.  The orientation can be defined by directors advected from a smooth initial orientation field, or it can be defined by the average orientation of an ensemble of initial orientations that can be quantified by the stretching eigenvectors.   The advected director field does not develop new topological singularities, so it will always remain smooth if it starts with a smooth initial condition. However, the stretching eigenvector field does develop new topological singularities; in this chaotic flow, the number of singularities increases exponentially.  Despite the very different topology, the two orientation fields converge at long times indicating that the topological singularities are not the key coherent structures in these orientation fields.  Instead, emergent scar lines dominate both orientation fields at long times. 

The mathematical foundations for the passive director problem are still much less developed than for the passive scalar problem~\cite{Aref:2017} or for the detection of Lagrangian coherent structures in velocity fields~\cite{Haller2015}.   The close connection between the phenomenological description of passive director fields developed here and work on strange eigenmodes and Lagrangian coherent structures suggests that significant progress on mathematical foundations of the passive director problem may be possible.  In particular, recent work that uses the eigenvectors of the Cauchy-Green strain tensors for coherent vortex detection~\cite{Karrasch:2014,Serra:2017} considers the same topological singularities that we study and should be able to be extended to the passive director problem.    

We also hope that future work can extend these insights to the case of turbulent flows that are pervasive in industrial and environmental fiber flows.  Emergent scar lines should appear in any flow with chaotic exponential stretching of material line elements. We expect that advected director fields in two-dimensional turbulence will be dominated by scar lines that are similar to the 2D chaotic flow case studied here. The Lagrangian coherent structures determined by fluid deformation in 2D chaotic flows are similar to those found in turbulent flows~\cite{Mathur:2007,Twardos:2008}.  It should be possible to determine the age of emergent scar lines in 2D turbulence and select the age most relevant to observations at a given resolution. In 3D turbulence, the situation is less clear.  Methods for detection of Lagrangian coherent structures in 3D turbulence using measures of fluid stretching have shown promise~\cite{Green:2007}.  Analysis of fluid stretching has been shown to be an effective way to understand the alignment of fibers and other non-spherical particles in 3D turbulent flows~\cite{Ni2014,Zhao:2016}.  Additional work is needed to determine whether the scar lines that dominate director orientation fields in 2D chaotic flow will appear in 3D flows.  

\section{Acknowledgments}

This work was supported by NSF Grant No.~DMR-1508575, Vetenskapsr\aa{}det Grant No.~2013-3992, and the Knut and Alice Wallenberg FoundationGrant No.~KAW 2014.0048.  We thank Stellan Ostlund, Nicholas Ouellette, and Michael Wilkinson for stimulating discussions.

\section{Appendix A}

To quantify fluid deformation, consider a point in the flow that is initially at $\mathbf{X}$ and is advected after time $\Delta t$ to $\mathbf{x}$.  The fluid deformation gradient is defined as $F_{ij}= \frac{\partial x_i}{\partial X_j}$. The deformation gradient includes both rotation and strain, $\mathbf{F}=\mathbf{VR}=\mathbf{RU}$, where $\mathbf{R}$ is the rotational tensor and $\mathbf{V}$ and $\mathbf{U}$ are the left and right stretch tensors respectively \cite{Ni2014}. It is convenient to extract only the strain contribution using the Cauchy-Green strain tensors.  The left Cauchy-Green strain tensor, $\mathbf{C}^{(L)}=\mathbf{F}\mathbf{F}^{T}=\mathbf{VR}\mathbf{R}^{T} \mathbf{V}^{T}=\mathbf{VV}$, has eigenvectors along the principle axes of the ellipse formed after the fluid element is deformed over $\Delta t$. The right Cauchy-Green strain tensor, $\mathbf{C}^{R}=\mathbf{F}^{T}\mathbf{F}=\mathbf{U}^{T} \mathbf{R}^{T} \mathbf{RU}=\mathbf{UU}$, has eigenvectors along the initial direction that will become the principal axes after deformation. Thus the field formed by the eigenvector of the Left Cauchy-Green strain tensor gives the preferred direction toward which a fiber at that location will have rotated due to the fluid deformation. Both the right and left Cauchy-Green strain tensors have the same eigenvalues, $
\Lambda_1$ and $\Lambda_2$ with $\Lambda_1$ traditionally chosen to be the maximum (extensional) eigenvector.   The square root of the maximum eigenvalue gives the stretching the fluid element experiences, defined as the ratio of the final major axis of the elliptical fluid element divided by the initial diameter.   The finite-time Lyapunov exponents are defined by $\lambda_i = \frac{1}{ t} \ln{\sqrt{\Lambda_i}}$.

An alternative way to express the final preferred orientation is to calculate the eigenvectors of a tensor order parameter. The tensor order parameter that is widely used in the study of liquid crystals is
\begin{equation}
I_{ij}=\frac{1}{2\pi}\int^{2\pi}_{0} d\theta \; P(\hat p(\theta), \mathbf{r}, t)\,(\hat p_{i} \hat p_{j}-\frac{1}{3}\delta_{ij}).
\end{equation}
Wilkinson \textit{et al.}~\cite{Wilkinson2010} used an order parameter without the isotropic term.
For initially uniform $P(\hat p(\theta), \mathbf{r}, t=0)$, both of these tensor order parameters have the same eigenvectors as the left Cauchy-Green strain tensor.

%

\end{document}